\documentclass[letterpaper, 10 pt, conference]{ieeeconf}  

\IEEEoverridecommandlockouts                              
\usepackage[utf8]{inputenc}
\usepackage{graphicx}

\usepackage{epsfig}
\usepackage{epstopdf}
\usepackage{amsmath}

\usepackage{caption}
\usepackage{subcaption}
\usepackage{multicol, blindtext}
\usepackage{url}
\usepackage{hyperref}
\usepackage[capitalize]{cleveref}
 \usepackage[ruled]{algorithm2e}
\usepackage{algorithmicx}

 \usepackage{pgfplots}
  \pgfplotsset{compat=newest}

  \usetikzlibrary{plotmarks}
  \usetikzlibrary{arrows.meta}
  \usepgfplotslibrary{patchplots}
  \usepackage{grffile}

\title{\LARGE \bf
En Route Path-planning for Partially Occupied Vehicles in Ride-pooling Systems
}

\author{Pengbo Zhu$^{1}$, Giancarlo Ferrari-Trecate$^{2}$, Nikolas Geroliminis$^{1}$
\thanks{*Research supported by NCCR Automation, a National Centre of Competence in Research, funded by the Swiss National Science Foundation (grant number 51NF40\_180545)}
\thanks{$^{1}$   \'Ecole Polytechnique F\'ed\'erale de Lausanne, Urban Transport Systems Laboratory, 1015 Lausanne, Switzerland, {\tt\small pengbo.zhu@epfl.ch,  nikolas.geroliminis@epfl.ch}}
\thanks{$^{2}$ \'Ecole Polytechnique F\'ed\'erale de Lausanne, Dependable Control and Decision Group, 1015 Lausanne, Switzerland, {\tt\small giancarlo.ferraritrecate@epfl.ch}}
}

\begin{document}

\maketitle
\thispagestyle{empty}
\pagestyle{empty}

\begin{abstract}
Ride-pooling services, such as UberPool and Lyft Shared Saver, enable a single vehicle to serve multiple customers within one shared trip. Efficient path-planning algorithms are crucial for improving the performance of such systems. For partially occupied vehicles with available capacity, we introduce a novel routing algorithm designed to maximize the likelihood of picking up additional passengers while serving the current passengers to their destination. 
Unlike traditional methods that group passengers and vehicles based on predefined time windows, our algorithm allows for immediate responses to passenger requests. Our approach optimizes travel time while dynamically considering passenger demand and coordinating with other vehicles. Formulated as an integer linear programming (ILP) problem, our method is computationally efficient and suitable for real-time applications. Simulation results demonstrate that our proposed method can significantly enhance service quality.

\end{abstract}

\section{Introduction}\label{sec: Intro}
Urban mobility demand is rising rapidly due to the expansion of metropolitan areas and increasing population. Compared to public transportation, the ride-sourcing service offers a more convenient and privacy-preserving mobility alternative, by efficiently connecting drivers and passengers through mobile internet. Moreover, as a part of the sharing economy, it enhances vehicle utilization \cite{Ke2020} and helps reduce air pollution. However, this expanding market also brings challenges, such as developing efficient car-passenger matching \cite{Pelzer2015}, optimizing routing \cite{AlonsoMora2017B}, and real-time fleet operations \cite{Alex2019, Zhu2024}.

In addition to the traditional service model, where one vehicle serves one ride, ride-sourcing providers are introducing ride-pooling services, such as UberPool and Lyft Shared Saver, which allow two or more passengers to share their trips in a single vehicle \cite{Storch2021, Ke2021}. It improves urban mobility accessibility by utilizing available seats, not only increasing driver profits but also helping alleviate traffic congestion. What's more, passengers benefit from discounted trip costs when sharing rides \cite{Zhang2023}, making it a win-win-win situation for the urban environment, drivers, and passengers.

One critical challenge in ride pooling services is the routing problem. 
Many studies have been conducted to find route planning for vehicles and passengers that optimizes a prescribed objective. For example, an online ride-sharing system was designed in \cite{Cici2015} to maximize the number of matched requests, where the passengers send their requests for a ride in advance. Besides, \cite{Asghari2016} focused on maximizing the total revenue of the system, while \cite{HOSNI2014} formulated a mixed-integer programming approach to maximize driver profit and minimize passenger travel time and costs.
A unified approach for route planning was proposed in \cite{Tong2018} for shared mobility, which optimized the sequence of pickups and dropoffs while managing conflicting objectives. Most of these studies, however, assume static conditions where the travel requests are fully known in advance. In practical applications, dynamic route planning is essential \cite{FURUHATA2013}, due to the time-varying traffic conditions and the stochastic nature of passenger demand. To address this, an Approximate Dynamic Programming method is introduced in \cite{Shah2020} to enable real-time routing that can respond effectively to time-varying demand. 

Different from conventional route planning which typically suggests the shortest path for drivers, the ride-pooling scenario allows a certain detour within the threshold in exchange for reduced fares for passengers. In \cite{Wang2019}, it was claimed that a high-quality planning path crowded with potential passengers is more valuable than a low-quality path for both drivers and passengers.
A detour-planning algorithm was also examined in crowd-sourcing systems by \cite{Chen2013}, where workers were encouraged to take detours to complete additional tasks for profit without exceeding their original deadlines. However, existing research lacks fine-grained detour path planning algorithms for ride-pooling services, especially in dynamic, real-time routing contexts.

In this work, we formulate the detour planning problem as an Arc Orienteering Problem (AOP) \cite{SOUFFRIAU2011}, a routing category focused on determining a sequence of edges to visit that maximizes the total collected score or profit within a specified time or resource budget. More specifically, we address path planning for partially occupied taxis, designing detailed routes from their current positions to the current passenger's destination with the goal of maximizing the likelihood of picking up additional passengers en route. At the same time, this approach also facilitates coordination with other vehicles in the urban area. Rather than batching customers and vehicles based on a predetermined time window, our algorithm enables real-time responses to passenger requests.

The remainder of this paper is organized as follows:
In \cref{sec: motivation}, the motivation for the path-planning problem for partially occupied vehicles is provided. In \cref{sec: Modeling}, we model the matching probability considering two key factors: (1) Attractiveness, regarding passenger demand, and (2) Repulsiveness, related to the influence of other vehicles. We then approximate the objective function in linear form and formulate the path-planning task as an integer linear programming problem in \cref{sec: Problem Formulation}. The proposed algorithm is tested on a ride-sharing simulator using the real city road network of Shenzhen, China, in \cref{sec: Case Study}. Finally, we conclude the work and present future research directions in \cref{sec: Conclusion}.

\section{Problem Statement and Modeling} \label{sec: Preliminary}

\subsection{Motivation}\label{sec: motivation}
\begin{figure}[ht]
    \centering
    \includegraphics[width=1\linewidth]{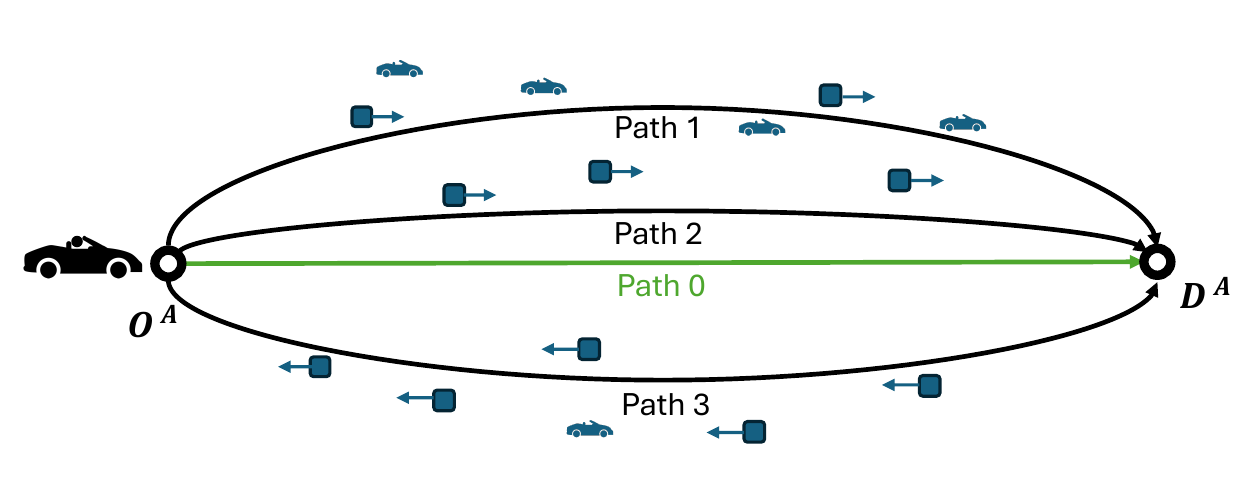}
    \caption{{\small Four different path choices for a partially occupied vehicle. The shortest path, labeled ‘Path 0’, is shown in green. Waiting passengers are indicated by square markers, with arrows pointing to their intended travel directions. The arc lengths represent the lengths of each path.}}
    \label{fig:examplePath}
    \vspace{-0.2cm}
\end{figure}
Assume that taxis have a capacity of two, meaning they can serve up to two passengers with separate origins and destinations on one trip. A partially occupied vehicle has already picked up one passenger at their origin $O^A$ and is en route to the passenger’s destination $D^A$. There are multiple candidate paths as shown in \cref{fig:examplePath}. Path 0, the shortest route between $O^A$ and $D^A$ in terms of distance, has no waiting passengers along it. Thus, if the vehicle follows Path 0, it will complete a solo trip without picking up additional passengers. Path 3, although populated with waiting passengers, primarily has passengers traveling in the opposite direction of the current trip, from $O^A$ to $D^A$, making it less viable for pooling. There are waiting passengers whose trips are compatible with the current one near Paths 1 and 2, but Path 1 has more competing vehicles for matching. Therefore, even though Path 2 involves a detour, making it longer than the shortest route, it is considered a more preferable option due to the greater likelihood of successfully pooling another passenger along this path.

This example demonstrates that, in a ride-pooling context, the shortest path is not necessarily the optimal choice. Instead, an efficient path-planning algorithm for partially occupied vehicles is essential to improve operational efficiency and service quality. 
In this work, we focus on the path-planning problem that finds an efficient route from the origin to the destination of the first passenger, maximizing the likelihood of picking up an additional passenger en route. At the same time, the planned routes should not exceed the acceptable detour distance threshold of passengers who share a ride.

Based on the observations above, an effective route should consider the following factors:
\begin{itemize}
    \item \textbf{Attractiveness}: Passenger demand. By routing through areas with high concentrations of requests with similar travel directions, a vehicle increases its likelihood of matching with a second passenger.
    
    \item \textbf{Repulsiveness}: Competition from other vehicles. With centralized control and operation, vehicle positions and planned routes are shared across the entire fleet. Coordinating these vehicles enables route planning that avoids oversupply or undersupply in urban areas.
\end{itemize}

In the following subsection, a matching probability model is developed to quantify the likelihood of a partially occupied vehicle picking up a second passenger en route to the current passenger's destination. This model is used as the foundation for our path-planning algorithm, which aims to maximize this likelihood by considering the aforementioned two factors.

\subsection{Trip-specific Matching Probability Modeling}\label{sec: Modeling}
The road network is represented by a graph $G = (Q, E, \omega)$, where $Q$ is the set of nodes (road intersections), $E$ is the set of edges (road segments) and $\omega$ is the length of each road segment accordingly. A passenger request can be described through a pair $(O^A, D^A)\in Q\times Q$, where $O^A$ and $D^A$ represent the origin and destination nodes, respectively, with $O^A \neq D^A$. 

Inspired by the matching function considered in \cite{Buchholz2021} and the empirical law of pooling-matching probability found in \cite{Ke2021}, we formulate the probability of a match for a partially occupied vehicle at node $i$, given the current trip $(O^A, D^A)$, as follows
\begin{equation}\label{model: match}
    p^t_{node}(i, O^A, D^A) = 1- \zeta exp \left(- \frac{\lambda_{Att}^t(i, O^A, D^A)}{\eta n^t(i)}\right),
\end{equation}
where $\zeta \in (0, 1]$ is a parameter related to the matching radius $d^t_m$ and the distributions of passengers and vehicles. The matching radius defines the area within which the vehicle searches for potential passengers. 
 A larger matching radius increases the potential for finding a match but may also lead to a longer distance to pick up the passenger, while a smaller radius limits the search area.
$\eta$ is a region-specific positive parameter related to the difficulty of searching available vehicles, influenced by factors such as the network structure. 

Some studies have been devoted to modeling the matching probability at the network level (see \cite{Yan2020, Ke2021}). However, these models are not suitable for our focus on trip-specific pool-matching probabilities. To account for pooling constraints, such as acceptable detour distances for individual passengers, we define the expected compatible passenger demand at node $i$ for a partially occupied vehicle currently on trip $(O^A, D^A)$ as
\begin{equation}\label{attractiveness}
    \lambda_{Att}^t(i, O^A, D^A) = \sum_{j\in Q}\lambda^t(i,j)  \frac{\mathcal{L}_s(O^A, D^A)+ \mathcal{L}_s( i,j)}{2\mathcal{L}_p (O^A, D^A, i, j)},
\end{equation}
where $\lambda^t(i,j)$ is the Poisson arrival rate of customers requesting trips from node $i$ to node $j$ at time $t$, estimated from historical data.
$\mathcal{L}_s(a,b)$ is a function that returns the traveling distance if a vehicle travels directly from node $a$ to $b$ following the shortest path calculated by Floyd-Warshall Algorithm \cite{Floyd1962}. $\mathcal{L}_p(O^A, D^A, i, j)$ returns the minimal combined traveling distance when passenger $A$ pools a ride with another passenger requesting a trip from $i$ to $j$.  

In this work, we consider two possible shared traveling sequences for ride-pooling: $O^A \rightarrow O^B \rightarrow D^A \rightarrow D^B$ (first pick-up, first drop-off) and $O^A \rightarrow O^B \rightarrow D^B \rightarrow D^A$ (first pick-up, last drop-off). Accordingly, $\mathcal{L}_p(O^A, D^A, i, j)$ is defined as:
\begin{equation}\label{detourdist}
\begin{aligned}
\mathcal{L}_p(O^A, D^A, i, j) =
\min \{
 \mathcal{L}_s(O^A, i) + \mathcal{L}_s(i, j) + \mathcal{L}_s(j, D^A),& \\
 \mathcal{L}_s(O^A, i) + \mathcal{L}_s(i, D^A) + \mathcal{L}_s(D^A, j)\}&
\end{aligned}
\end{equation}

Compared with the traditional formulation $\lambda^t(i) = \sum_{j\in Q}\lambda^t(i,j)$ describing the Poisson arrival at node $i$ without considering pooling constraints (as used in \cite{Buchholz2021}), our formulation \eqref{attractiveness} includes the ratio {\large $ \frac{\mathcal{L}_s(i, j) + \mathcal{L}_s(O^A, D^A)}{2\mathcal{L}_p(O^A, D^A, i, j)}$} $\in (0, 1]$. This ratio serves as a weighting factor that quantifies the efficiency of pooling the two trips $(O^A, D^A)$ and $(i, j)$. If its value is close to 1, it indicates that the combined trip does not significantly increase the total travel distance compared to both passengers traveling separately, making the potential match more attractive; while a lower ratio suggests the pooling would involve a substantial detour. For example, when many trips from node $i$ are heading in the opposite direction of trip $(O^A, D^A)$, this results in a smaller ratio, decreasing the attractiveness of node $i$. Therefore, the entire expression in \eqref{attractiveness} quantifies the attractiveness of node $i$ for picking up a second passenger compatible with the current trip.

In addition, the repulsiveness due to the competition from other vehicles is considered in our model \cref{model: match}. Specifically, $n^t(i)$ represents the number of available vehicles surrounding node $i$ within a matching radius $d^t_m$ at time $t$, calculated as
\begin{equation}
    n^t(i) = n^t_{empty}(i) + n^t_{drop}(i) + 0.5 n^t_{partial}(i),
\end{equation}
where captures the effective number of available vehicles near node $i$, accounting for different levels of availability. Empty vehicles, represented by $n^t_{empty}(i)$, are fully available for new passenger requests. Vehicles in the process of dropping off their final passenger, $n^t_{drop}(i)$, will soon be available once they reach their destinations.
Partially occupied vehicles, denoted by $n^t_{partial}(i)$, are assigned a multiplier of $0.5$, as they can only accept one additional passenger, i.e., half of their capacity. 

\subsection{Likelihood of getting a second passenger on edge}
Assuming that the start and end nodes contribute equally to a potential match on a road segment, the probability of a match event occurring on the edge $(i,j) \in E$, starting at node $i$ and ending at $j$, is defined as:
\begin{equation}
\begin{aligned}
p^t_{edge}(i, j, O^A, D^A) = (p^t_{node}&(i, O^A,  D^A)\\
&+ p^t_{node}(j, O^A, D^A))/2.
\end{aligned}
\end{equation}

The travel time on edge $(i, j)$ can be estimated using measurement methods such as infrared loop detector, historical data, or real-time vehicle GPS trajectories \cite{Wang2014}, denoted as $\hat{T}(i, j)$. By assuming that match opportunities occur independently over time, the probability of not obtaining a match during the travel time $\hat{T}(i, j)$ is $(1 - p^t_{edge}(i, j, O^A, D^A))^{\hat{T}(i,j)}$. Therefore, the probability of a partially occupied vehicle picking up a second passenger while traveling along edge $(i, j)$ is given by
\begin{equation}\label{eq: edge_prob}
\mathcal{P}^t_{ij} = 1 - (1 - p^t_{edge}(i, j, O^A, D^A))^{\hat{T}(i,j)}.
\end{equation}

\section{En route Path-planning Algorithm}\label{sec: Problem Formulation} 
\subsection{Objective Function}
A path is planned for a partially occupied vehicle from a given start node (origin of current passenger $O^A$) to an end node (destination of current passenger $D^A$) that maximizes the likelihood of picking up a second passenger en route. We define the decision variable $x_{ij}$ as follows:
\begin{equation}
    x_{ij} = 
\begin{cases} 
1 & \text{if edge $(i, j) \in E$ is part of the path,} \\
0 & \text{otherwise.}
\end{cases}
\end{equation}

The overall probability of not picking up any passenger along the planned route, which includes a sequence of edges, can be determined by multiplying the probability of not picking up a second passenger on each individual edge. 
Thus, our objective can be expressed as
\begin{equation}\label{orginalObj}
    \max_{x_{ij}}  \left(J = 1 - \prod_{(i,j) \in E} \left(1 - \mathcal{P}_{ij}^t \cdot x_{ij}\right)\right)
\end{equation}

To simplify the objective function, we apply a logarithmic transformation to reformulate the maximization problem which gives:
\begin{equation}
    \begin{aligned}
    \max_{x_{ij}}  J &= \min_{x_{ij}}  \left( \log\left( 1 - J \right) \right) \\
    &= \min_{x_{ij}}  \left( \log\left( \prod_{(i,j) \in E} \left( 1 - \mathcal{P}_{ij}^t \cdot x_{ij} \right) \right) \right) \\
    &= \min_{x_{ij}}  \left( \sum_{(i,j) \in E} \log\left( 1 - \mathcal{P}_{ij}^t \cdot x_{ij} \right) \right).
\end{aligned}
\end{equation}

Given small values of $\mathcal{P}_{ij}^t \cdot x_{ij}$, we can use the approximation $\log(1 - y) \approx -y$ when $y$ is close to zero. 
Therefore, the objective function is approximated to a linear form as follows
\begin{equation}
    \max_{x_{ij}}   \sum_{(i,j) \in E} \mathcal{P}_{ij}^t \cdot x_{ij}.
\end{equation}

\subsection{Integer Linear Programming (ILP) for Partially Occupied Vehicle Path-Planning}

The en-route path planning problem can be formulated as the following ILP problem:
\begin{subequations}\label{ILP}
\begin{align}
& \max_{x_{ij}} \quad \sum_{(i, j) \in E} \mathcal{P}_{ij}^t \cdot x_{ij}, \\
\text{sub}&\text{ject} \text{ to}  \nonumber \\
& x_{ij} \in \{0, 1\}, \quad \forall (i, j) \in E,  \label{cons:Binary} \\
& \sum_{j: (O^A, j) \in E} x_{O^A, j} = 1,  \sum_{i: (i, D^A) \in E} x_{i, D^A} = 1, \label{cons:StartEnd} 
\\
& \sum_{j: (i,j) \in E}x_{ij}- \sum_{k: (k,i) \in E}x_{ki} = 0, \forall i \in N \setminus \{O^A, D^A\}, \label{cons:flowcontinuity} \\
& \sum_{j: (i,j) \in E} x_{ij} \leq 1, \quad \sum_{k: (k,i) \in E} x_{ki} \leq 1, \quad \forall i \in N, \label{cons:visitOnce} \\
& \sum_{(i, j) \in E} \omega_{ij} \cdot x_{ij} \leq \alpha \cdot \mathcal{L}(O^A, D^A), \quad \alpha \geq 1, \label{cons:MaxDetour}
\end{align}
\end{subequations}
where $\mathcal{P}_{ij}^t$ is the probability of getting a second passenger while traveling along edge $(i,j)$ at time $t$, as described in \cref{eq: edge_prob}, and $\omega_{ij}$ represents the length of edge $(i, j)$.
Constraint~\eqref{cons:Binary} enforces the binary nature of the decision variables.  Constraint \eqref{cons:StartEnd} guarantees that the route begins at node $O^A$ (the origin of the current passenger) and ends at node $D^A$ (the destination of the current passenger). Constraint~\eqref{cons:flowcontinuity} ensures flow continuity at all nodes except the start and end nodes, maintaining a continuous path. Constraint~\eqref{cons:visitOnce} ensures that each node is visited at most once, preventing cycles in the path. Finally, the total length of the planned route is limited in ~\eqref{cons:MaxDetour} to be no more than $\alpha$ times the length of the shortest path $\mathcal{L}_s(O^A, D^A)$, i.e., the maximum detour distance. Solving this ILP problem provides the optimal set of edges, which maximizes the likelihood of picking up an additional passenger along it while adhering to constraints. The route $r_{current}$ is then constructed by sequentially connecting these selected edges, forming a continuous path from the origin $O^A$ to the destination $D^A$. The procedure for implementing this algorithm is summarized in Algorithm \ref{alg:path}.

\begin{algorithm}[htb] 
\SetAlgoLined
When a vehicle picks up its first passenger:\\
        \textbf{Input:} Origin $O^A$ and destination $D^A$ of the first passenger; passenger arrival rates $\lambda_{OD}^t(i, j)$, vehicle speed $v^t$.\\
        
        \textbf{Communicate:} Obtain from the central dispatch system:\\
        1) Positions of all other empty vehicles.\\
        2) Destinations of vehicles currently dropping off their last passenger.\\
        3) Planned routes of other partially occupied vehicles: $\mathcal{R} = \{ r_i \}, \quad i = 1, 2, \dots, n^t_{partial}$.\\
        
        \textbf{Compute:}\\
        1) Calculate the match probabilities $\mathcal{P}_{ij}^t$ along each edge using \cref{eq: edge_prob}.\\
        2) Solve (\ref{ILP}) to determine the optimal set of edges forming the route $r_{current}$.\\

        \textbf{Move:} Move towards destination $D^A$ following the planned route $r_{current}$.\\
        
        \textbf{Update:} Augment the set of planned routes:
        $\mathcal{R} \leftarrow \mathcal{R} \cup \{ r_{current} \}$.
        \\$\quad n^t_{\text{partial}} \leftarrow n^t_{\text{partial}} + 1$
 \caption{Implementation of ILP for En-route Path Planning for Partially Occupied Vehicles}\label{alg:path}
\end{algorithm}

\section{Case Study}\label{sec: Case Study}
\subsection{Matching scheme}\label{MatchingScheme}
A basic matching policy is implemented in the simulation, following a “first-come, first-served” approach. When a passenger request is issued, the platform searches for the nearest empty vehicles to the passenger’s origin. If the closest empty vehicle can reach the passenger within a predetermined waiting time threshold, $T_w$, the request is assigned to this vehicle. If no empty vehicle is available and the passenger is open to ride-pooling, the platform then searches for the nearest partially occupied vehicle whose current passenger is also willing to share the ride. A match will happen if the detour distance for each passenger does not exceed $\alpha$ times the length of their shortest path. In this simulation, cancellations after matching are not permitted. However, if no available vehicles are found within a maximum period, $T_m$—the duration the request remains in the matching pool—the passenger will cancel the request.
    \vspace{-0.1cm}
\subsection{Experimental setup}
    \vspace{-0.4cm}
\begin{figure}[!ht]
    \centering
    \includegraphics[width=0.99\linewidth]{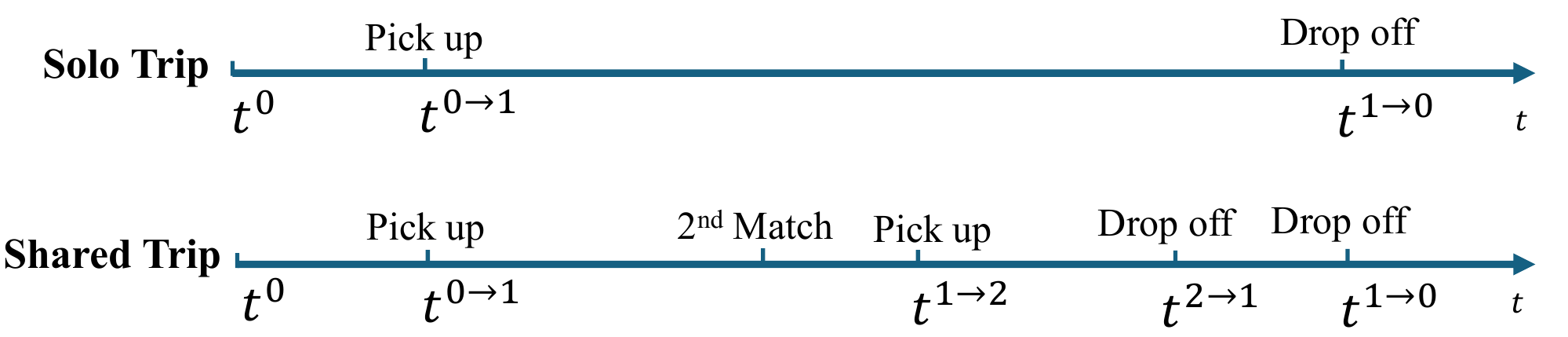}
    \caption{{\small Evolution of the number of passengers}}
    \label{fig:solo_shared}
        \vspace{-0.3cm}
\end{figure}

In a ride-sharing scenario, a taxi can serve multiple passengers simultaneously. Specifically, we focus on the common case where two ride requests are shared. Initially, at time $t^0$, the vehicle is empty and idle. The evolution of the number of passengers in the taxi over time is illustrated in \cref{fig:solo_shared}.

\begin{table*}[!ht]
 \vspace{2ex}
  \begin{center}
    \caption{Performance metrics}

    \begin{tabular}{|c||c|c|c|c|c|} 
    \hline
              \textbf{   }&\textbf{Answer rate} & \textbf{Av. waiting time} & \textbf{No. of shared orders} & \textbf{Total shared dist}  &\textbf{Total empty dist}\\
              \textbf{  }&\textbf{ ($\%$)} & \textbf{($s$)} 
              &\textbf{} 
              &\textbf{($km$)}  & \textbf{$(km)$}\\

      \hline

Proposed     &81.4  &151.3&656 &865.6  &1772.8 \\

\hline
Shortest  &80.0 &153.8 &614 &771.9&1845.1\\
\hline 

No share      &72.2 &162.3&0 &-  &1917.0\\
\hline \hline
No share ($fleet +20\%$) &80.4 &148.8&0 &-  &3026.3\\
      \hline
    \end{tabular}\label{table: path}
  \end{center}
   \vspace{-0.5cm}
\end{table*}

\begin{figure*}[ht]
    \centering
    \begin{subfigure}[b]{0.47\textwidth}
        \includegraphics[width=\textwidth]{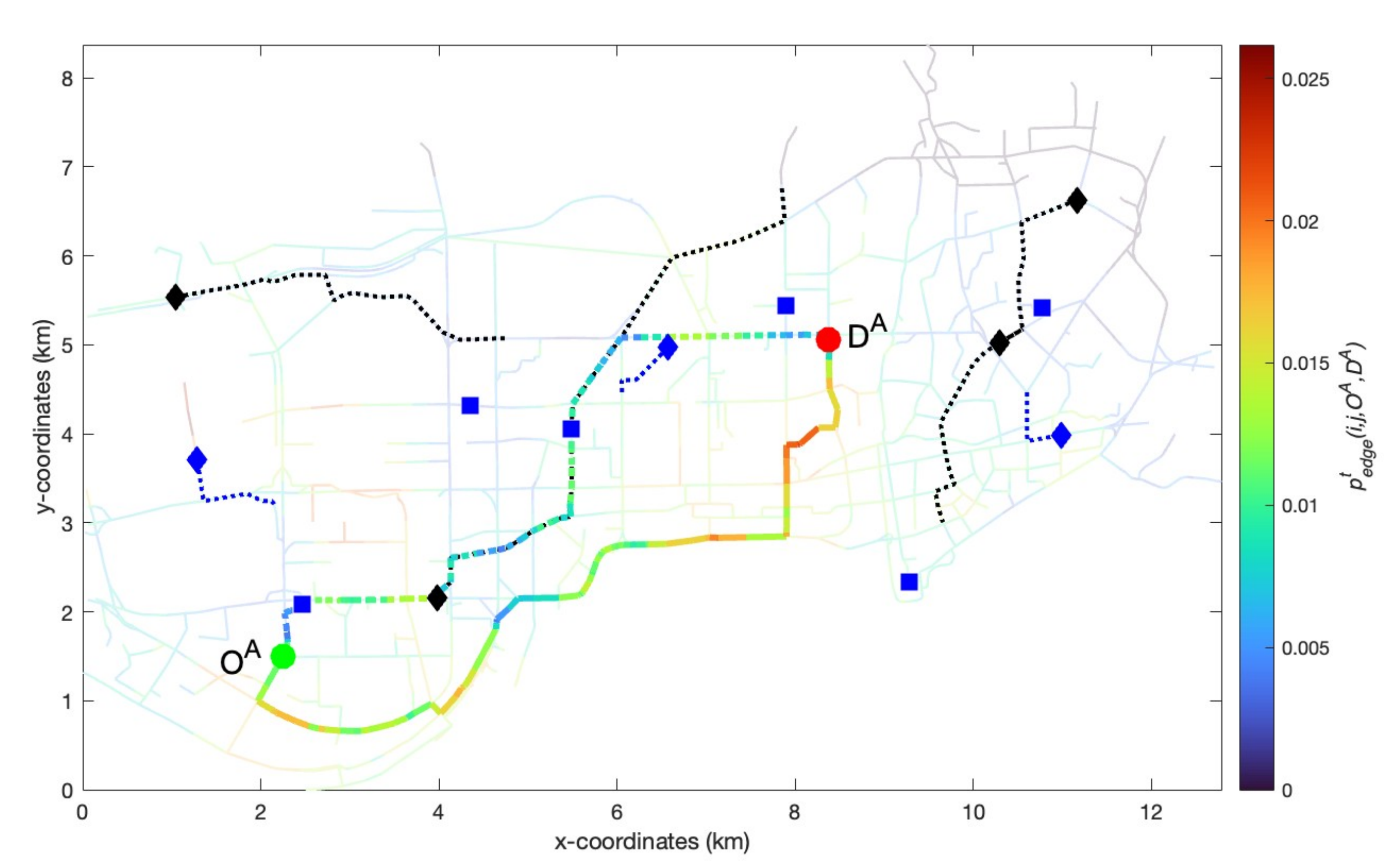}
        \caption{{\small Low-demand hour. Length of planned path: 10.39 km.}}
        \label{fig:Hour1}
    \end{subfigure}
    \begin{subfigure}[b]{0.47\textwidth}
        \includegraphics[width=\textwidth]{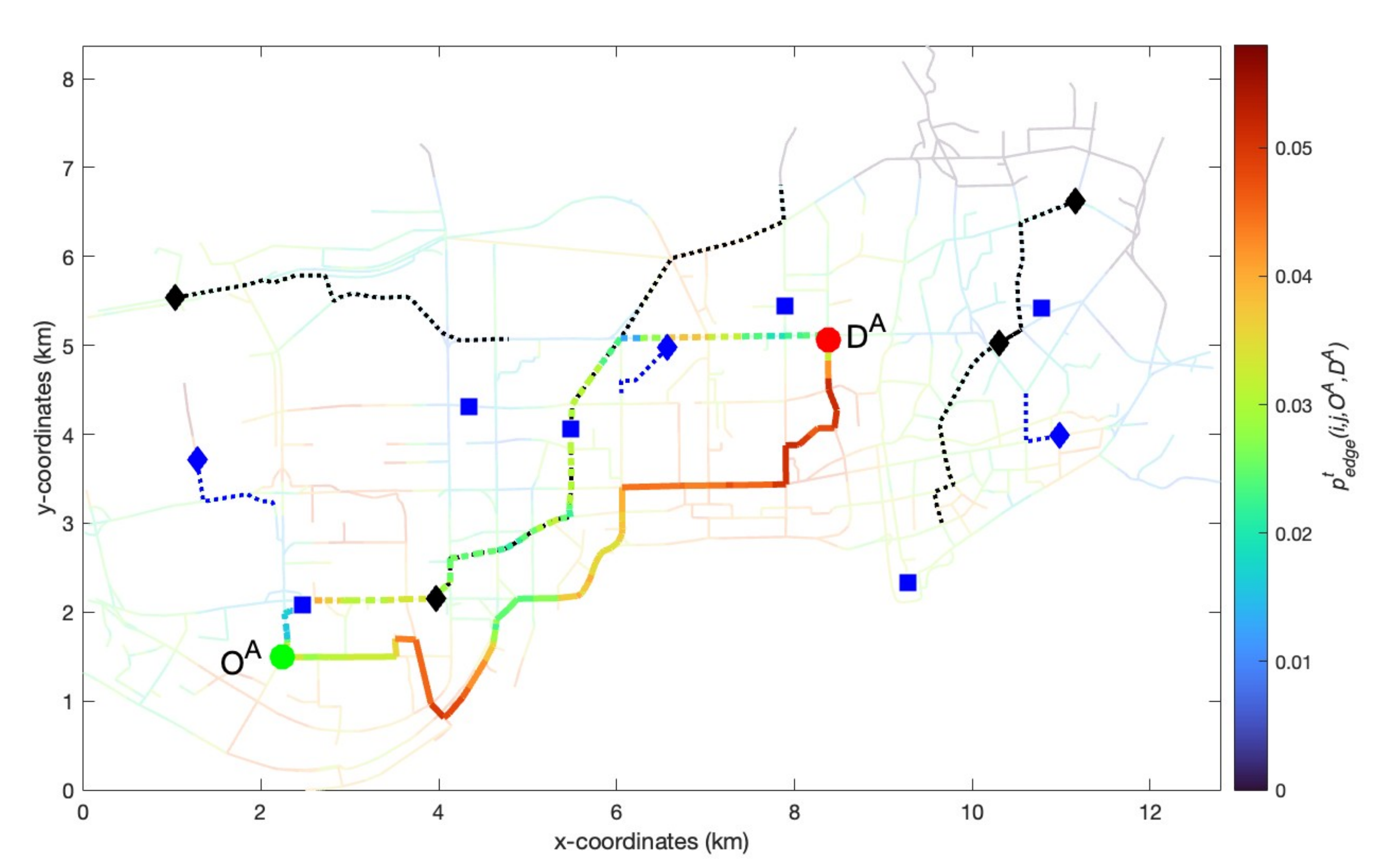}
        \caption{{\small High-demand hour. Length of planned path: 10.29 km.}}
        \label{fig:Hour2}
    \end{subfigure}
    \caption{{\small Illustration of the paths for a partially occupied vehicle traveling from the origin $O^A$ of its first passenger to the destination $D^A$ under different demand levels: (a) low-demand hour and (b) high-demand hour.
    A heat map is superimposed over the road network, representing the matching probability $p^t_{edge}(i, j, O^A, D^A)$ of each edge $(i, j)$. The shortest path between $O^A$ and $D^A$ is depicted as a dashed line, with edges colored according to the matching probability. Its length is $8.83$ km. The routes planned by our proposed method are shown as solid colored lines. Empty vehicles are indicated by blue square-shaped markers. The current positions and planned routes of other partially occupied vehicles are represented by black diamond-shaped markers and black dotted lines. Vehicles dropping off their last passengers are shown as blue diamonds.}
    }
    \label{fig:path_simulator}
\vspace{-0.6cm}
\end{figure*}

At time $t^{0 \rightarrow 1}$, the vehicle picks up its first passenger (Passenger $A$) at origin $O^A$. It then follows a route to destination $D^A$, chosen either as the shortest path or via our proposed method, depending on the operational strategy. En route, the vehicle may be matched with a second passenger (Passenger $B$) if the sharing constraints in \cref{MatchingScheme} are met.
When a match with Passenger $B$ occurs, the vehicle temporarily halts its planned route to $D^A$ and instead takes the shortest path to pick up Passenger $B$ at their origin $O^B$ at time $t^{1 \rightarrow 2}$. After picking up Passenger $B$, the vehicle proceeds to drop off passengers.
The drop-off sequence is chosen based on the total travel distance, selecting the shorter of the two possible routes: $O^B \rightarrow D^B \rightarrow D^A$ or $O^B \rightarrow D^A \rightarrow D^B$. Once both passengers have been dropped off, the vehicle becomes idle again at time $t^{1 \rightarrow 0}$. 
Conversely, if the vehicle fails to match with a second passenger or if the passengers are unwilling to share, the trip remains a solo journey. In this case, the vehicle becomes empty again after dropping off Passenger $A$ at $D^A$.

\vspace{-0.05cm}
The proposed algorithm was tested using a shared-mobility simulator \cite{Caio2021} based on the urban road network of the Luohu and Futian districts in Shenzhen, China, including 1,858 intersections and 2,013 bidirectional road segments. We consider a three-hour testing simulation, where 400 requests per hour are issued in the first and third hours, i.e., low-demand hours and 800 requests per hour are issued in the second hour, i.e., the high-demand hour. The fleet size is 100, and their initial locations are randomly generated and evenly distributed over the road network. The passenger detour distance threshold is set to $\alpha = 1.2$, allowing detours up to $20\%$ longer than the shortest path. The passenger waiting and matching time threshold are $T_w = 5 \text{min}$ and $T_m = 1 \text{min}$, respectively.
In our method, other parameters such as $\zeta$ and $\eta$ are tuned via grid search, evaluating performance across combinations to maximize the answer rate.  In the case study, we apply uniform values of $\zeta = 1$ and $\eta = 0.001$ across all nodes, as the matching scheme runs on a centralized cloud platform.
For the sake of brevity, we assume a constant traveling speed when vehicles travel along a planned route. The travel speed is given by a macroscopic fundamental diagram (MFD, see \cite{Caio2021}), which describes the relationship between the accumulation of vehicles (i.e., the total number of vehicles) and the space-mean speed $v^t$. Then, the travel time for edge $(i, j)$ at time $t$, given its length $\omega_{ij}$ and speed $v^t$, can be calculated as $\hat{T}(i, j) = \frac{\omega_{ij}}{v^t}$. The corresponding matching radius is calculated by $d^t_m = T_w \cdot v^t \text{ km}$. The ILP (\ref{ILP}) is solved using Gurobi Optimizer (version 11.0.3).

\subsection{Results and Analysis}

To compare the service efficiency and fleet utilization, we consider the following performance metrics: \textbf{Answer rate}, which is the proportion of successfully completed orders relative to all passenger requests; \textbf{Av. waiting time}, which is the average time passengers spend waiting from the moment they issue a request until they are picked up, calculated as the total waiting time divided by the number of successfully completed orders; \textbf{No. of shared orders}, which is the total number of orders that are shared with another passenger; \textbf{Total shared distance}, measuring the cumulative distance traveled for shared rides; \textbf{Total empty distance}, which measures the distance traveled by vehicles when they are empty and cruising freely without passengers.

We compare our method with two other approaches. The first is the shortest path policy, referred to as “Shortest” in \cref{table: path}. In this approach, after picking up the first passenger, a taxi travels to the passenger’s destination using the shortest possible route in terms of travel distance, without considering traffic conditions. In both our proposed method and the “Shortest” policy, we assume that all taxis have a capacity of two and that all passengers are willing to share their trip. In contrast, the “No Share” method does not allow ride-pooling, meaning all passengers travel alone without sharing their trips with others.

With a capacity of two, ride-pooling methods (“Proposed” and “Shortest”) serve more requests and reduce passenger waiting times (\cref{table: path}).
Compared to the “Shortest” policy, our method results in a greater number of shared trips and a longer total shared distance, suggesting that it effectively plans routes to help taxis match with additional passengers for shared rides. Moreover, it shows a reduced total empty travel distance, reflecting improved vehicle utilization.
We further increased the fleet size for the “No Share” policy. Results indicate that achieving a comparable answer rate to ride-pooling methods requires a $20\%$ larger fleet under this policy.
However, the significantly higher empty travel distance will lead to unnecessary fuel consumption and traffic congestion in urban areas.

\cref{fig:path_simulator} provides a simulator snapshot where our proposed method plans the path for a taxi transporting a passenger from node $O^A$ to $D^A$. The planned route maintains distance from empty vehicles and those dropping off passengers, avoiding overlap with the planned routes of other partially occupied vehicles. Compared to the shortest path, our planned routes pass through edges of high attractiveness, as indicated by the warmer colors of the paths.

The planned paths in \cref{fig:Hour1} and \cref{fig:Hour2} remain within the $20\%$ detour threshold. Notably, the planned path during the low-demand hour is longer ($10.39$ km) than during the high-demand hour ($10.29$ km). It aligns with our expectations that, for the same origin and destination, less detour distance is needed during high-demand hours due to the higher probability of finding a second passenger. In contrast, low demand requires vehicles to travel farther to find additional passengers. This demonstrates that our method can adapt effectively to varying demand levels.

\section{Conclusion}\label{sec: Conclusion}
This paper presents a novel en-route path planning algorithm for partially occupied vehicles in ride-pooling systems. Instead of following the shortest path, our proposed method plans an efficient route that guides vehicles through high-demand areas while coordinating with other vehicles to maximize the likelihood of picking up a second passenger to share the current ride. We verify the performance of our method using an agent-based simulation of the urban network of Shenzhen, China, demonstrating that it enhances service quality and improves vehicle utilization.

The current work estimates passenger arrival rate using historical data. Future efforts will incorporate online prediction of stochastic demand. Accurate prediction of traveling speeds on each edge is also critical to be investigated. Furthermore, developing more efficient methods for hyperparameter tuning is another interesting direction for future research.
\vspace{-0.2cm}

\bibliographystyle{IEEEtran}
\bibliography{path_ref}

\begin{thebibliography}{10}
\providecommand{\url}[1]{#1}
\csname url@rmstyle\endcsname
\providecommand{\newblock}{\relax}
\providecommand{\bibinfo}[2]{#2}
\providecommand\BIBentrySTDinterwordspacing{\spaceskip=0pt\relax}
\providecommand\BIBentryALTinterwordstretchfactor{4}
\providecommand\BIBentryALTinterwordspacing{\spaceskip=\fontdimen2\font plus
\BIBentryALTinterwordstretchfactor\fontdimen3\font minus \fontdimen4\font\relax}
\providecommand\BIBforeignlanguage[2]{{%
\expandafter\ifx\csname l@#1\endcsname\relax
\typeout{** WARNING: IEEEtran.bst: No hyphenation pattern has been}%
\typeout{** loaded for the language `#1'. Using the pattern for}%
\typeout{** the default language instead.}%
\else
\language=\csname l@#1\endcsname
\fi
#2}}

\bibitem{Ke2020}
J.~Ke, H.~Yang, and Z.~Zheng, ``On ride-pooling and traffic congestion,'' \emph{Transportation Research Part B: Methodological}, vol. 142, pp. 213--231, 2020.

\bibitem{Pelzer2015}
D.~Pelzer, J.~Xiao, D.~Zehe, M.~H. Lees, A.~C. Knoll, and H.~Aydt, ``A partition-based match making algorithm for dynamic ridesharing,'' \emph{IEEE Transactions on Intelligent Transportation Systems}, vol.~16, no.~5, pp. 2587--2598, 2015.

\bibitem{AlonsoMora2017B}
J.~Alonso-Mora, A.~Wallar, and D.~Rus, ``Predictive routing for autonomous mobility-on-demand systems with ride-sharing,'' in \emph{2017 IEEE/RSJ International Conference on Intelligent Robots and Systems (IROS)}, 2017, pp. 3583--3590.

\bibitem{Alex2019}
A.~Wallar, J.~Alonso-Mora, and D.~Rus, ``Optimizing vehicle distributions and fleet sizes for shared mobility-on-demand,'' in \emph{2019 International Conference on Robotics and Automation (ICRA)}, 2019, pp. 3853--3859.

\bibitem{Zhu2024}
P.~Zhu, G.~Ferrari-Trecate, and N.~Geroliminis, ``Hierarchical control for vehicle repositioning in autonomous mobility-on-demand systems,'' \emph{IEEE Transactions on Control Systems Technology}, pp. 1--14, 2024.

\bibitem{Storch2021}
D.-M. Storch, M.~Timme, and M.~Schröder, ``Incentive-driven transition to high ride-sharing adoption,'' \emph{Nature Communications}, vol.~12, 06 2021.

\bibitem{Ke2021}
J.~Ke, Z.~Zheng, H.~Yang, and J.~Ye, ``Data-driven analysis on matching probability, routing distance and detour distance in ride-pooling services,'' \emph{Transportation Research Part C: Emerging Technologies}, vol. 124, p. 102922, 2021.

\bibitem{Zhang2023}
W.~Zhang, A.~Jacquillat, K.~Wang, and S.~Wang, ``Routing optimization with vehicle–customer coordination,'' \emph{Management Science}, vol.~69, no.~11, pp. 6876--6897, 2023.

\bibitem{Cici2015}
B.~Cici, A.~Markopoulou, and N.~Laoutaris, ``Designing an on-line ride-sharing system,'' in \emph{Proceedings of the 23rd SIGSPATIAL International Conference on Advances in Geographic Information Systems}, ser. SIGSPATIAL '15.\hskip 1em plus 0.5em minus 0.4em\relax New York, NY, USA: Association for Computing Machinery, 2015.

\bibitem{Asghari2016}
M.~Asghari, D.~Deng, C.~Shahabi, U.~Demiryurek, and Y.~Li, ``Price-aware real-time ride-sharing at scale: an auction-based approach,'' in \emph{Proceedings of the 24th ACM SIGSPATIAL International Conference on Advances in Geographic Information Systems}, 2016.

\bibitem{HOSNI2014}
H.~Hosni, J.~Naoum-Sawaya, and H.~Artail, ``The shared-taxi problem: Formulation and solution methods,'' \emph{Transportation Research Part B: Methodological}, vol.~70, pp. 303--318, 2014.

\bibitem{Tong2018}
Y.~Tong, Y.~Zeng, Z.~Zhou, L.~Chen, J.~Ye, and K.~Xu, ``A unified approach to route planning for shared mobility,'' \emph{Proc. VLDB Endow.}, vol.~11, no.~11, p. 1633–1646, July 2018.

\bibitem{FURUHATA2013}
M.~Furuhata, M.~Dessouky, F.~Ordóñez, M.-E. Brunet, X.~Wang, and S.~Koenig, ``Ridesharing: The state-of-the-art and future directions,'' \emph{Transportation Research Part B: Methodological}, vol.~57, pp. 28--46, 2013.

\bibitem{Shah2020}
S.~Shah, M.~Lowalekar, and P.~Varakantham, ``Neural approximate dynamic programming for on-demand ride-pooling,'' \emph{Proceedings of the AAAI Conference on Artificial Intelligence}, vol.~34, pp. 507--515, 2020.

\bibitem{Wang2019}
B.~Wang, R.~Zhu, S.~Zhang, Z.~Zhao, X.~Yang, and G.~Wang, ``Ppvf: A novel framework for supporting path planning over carpooling,'' \emph{IEEE Access}, vol.~7, pp. 10\,627--10\,643, 2019.

\bibitem{Chen2013}
C.-C. Liao and C.-H. Hsu, ``A detour planning algorithm in crowdsourcing systems for multimedia content gathering,'' in \emph{Proceedings of the 5th Workshop on Mobile Video}, ser. MoVid '13.\hskip 1em plus 0.5em minus 0.4em\relax New York, NY, USA: Association for Computing Machinery, 2013, p. 55–60.

\bibitem{SOUFFRIAU2011}
W.~Souffriau, P.~Vansteenwegen, G.~{Vanden Berghe}, and D.~{Van Oudheusden}, ``The planning of cycle trips in the province of east flanders,'' \emph{Omega}, vol.~39, no.~2, pp. 209--213, 2011.

\bibitem{Buchholz2021}
N.~Buchholz, ``{Spatial Equilibrium, Search Frictions, and Dynamic Efficiency in the Taxi Industry},'' \emph{The Review of Economic Studies}, vol.~89, no.~2, pp. 556--591, 09 2021.

\bibitem{Yan2020}
C.~Yan, H.~Zhu, N.~Korolko, and D.~Woodard, ``Dynamic pricing and matching in ride-hailing platforms,'' \emph{Naval Research Logistics (NRL)}, vol.~67, no.~8, pp. 705--724, 2020.

\bibitem{Floyd1962}
R.~W. Floyd, ``Algorithm 97: Shortest path,'' \emph{Commun. ACM}, vol.~5, no.~6, p. 345, June 1962.

\bibitem{Wang2014}
Y.~Wang, Y.~Zheng, and Y.~Xue, ``Travel time estimation of a path using sparse trajectories,'' in \emph{Proceedings of the 20th ACM SIGKDD International Conference on Knowledge Discovery and Data Mining}, ser. KDD '14.\hskip 1em plus 0.5em minus 0.4em\relax New York, NY, USA: Association for Computing Machinery, 2014, p. 25–34.

\bibitem{Caio2021}
C.~V. Beojone and N.~Geroliminis, ``On the inefficiency of ride-sourcing services towards urban congestion,'' \emph{Transportation Research Part C: Emerging Technologies}, vol. 124, p. 102890, 2021.

\end{thebibliography}

\end{document}